\documentstyle[epsf]{menu95}
\def\shiftleft#1{#1\llap{#1\hskip 0.04em}}
\def\shiftdown#1{#1\llap{\lower.04ex\hbox{#1}}}
\def\thick#1{\shiftdown{\shiftleft{#1}}}
\def\b#1{\thick{\hbox{$#1$}}}
\newcommand{\be}{\begin{equation}}
\newcommand{\ee}{\end{equation}}
\newcommand{\bea}{\begin{eqnarray}}
\newcommand{\eea}{\end{eqnarray}}

\def\Young{\vbox
{\hrule\hbox{\vrule\hskip.340cm\vbox{\vskip.340cm}\vrule}\hrule}}
\def\Y#1{ \ifnum #1=1  \vcenter{\Young}                       \fi
\ifnum #1=2
\vcenter{\vbox{\hbox to 0.720cm{\Young\hss\Young}}}          \fi
\ifnum #1=11
\vcenter{\vbox to 0.720cm{\Young\vss\Young}}                 \fi
\ifnum #1=3
\vcenter{\vbox{\hbox to 1.05cm{\Young\hss\Young\hss\Young}}} \fi
\ifnum #1=21
\vcenter{\vbox to 0.720cm{\hbox to 0.715cm{\Young\hss\Young}
\vss\hbox to 0.715cm{\Young\hss}}} \fi
\ifnum #1=22
\vcenter{\vbox to 0.720cm{\hbox to 0.715cm{\Young\hss\Young}
\vss\hbox to 0.715cm{\Young\hss\Young\hss}}} \fi
\ifnum #1=111
\vcenter{\vbox to 1.05cm{\Young\vss\Young\vss\Young}}        \fi
\ifnum #1=4
\vcenter{\vbox{\hbox to 1.44cm{\Young\hss\Young
\hss\Young\hss\Young}}} \fi
\ifnum #1=5
\vcenter{\vbox{\hbox to 1.79cm{\Young\hss\Young\hss\Young
\hss\Young\hss\Young}}} \fi
\ifnum #1=6
\vcenter{\vbox{\hbox to 2.1cm{\Young\hss\Young\hss\Young
\hss\Young\hss\Young\hss\Young}}} \fi
\ifnum #1=51
\vcenter{\vbox to 0.720cm{\hbox to 1.79cm{\Young\hss\Young\hss\Young
\hss\Young\hss\Young}\vss\hbox to 1.79cm{\Young\hss}}} \fi
\ifnum #1=42
\vcenter{\vbox to 0.720cm{\hbox to 1.44cm{\Young\hss\Young\hss\Young
\hss\Young}\vss\hbox to 0.715cm{\Young\hss\Young\hss}}} \fi
\ifnum #1=33
\vcenter{\vbox to 0.720cm{\hbox to 1.05cm{\Young\hss\Young\hss\Young}
\vss\hbox to 1.05cm{\Young\hss\Young\hss\Young}
}}      \fi
\ifnum #1=31
\vcenter{\vbox to 0.720cm{\hbox to 1.05cm{\Young\hss\Young\hss\Young}
\vss\hbox to 1.05cm{\Young\hss}
}}      \fi
\ifnum #1=222
\vcenter{\vbox to 1.05cm{\hbox to 0.715cm{\Young\hss\Young}
\vss\hbox to 0.715cm{\Young\hss\Young}
\vss\hbox to 0.715cm{\Young\hss\Young}}}\fi
\ifnum #1=211
\vcenter{\vbox to 1.05cm{\hbox to 0.715cm{\Young\hss\Young}
\vss\hbox to 0.715cm{\Young\hss}
\vss\hbox to 0.715cm{\Young\hss}}}\fi
\ifnum #1=111111
\vcenter{\vbox to 2.1cm{\Young\vss\Young\vss\Young
\vss\Young\vss\Young\vss\Young}}\fi
\ifnum #1=1111
\vcenter{\vbox to 2.1cm{\Young\vss\Young\vss\Young
\vss\Young\vss}}\fi}
\def\young{\vbox
{\hrule\hbox{\vrule\hskip.190cm\vbox{\vskip.190cm}\vrule}\hrule}}
\def\y#1{ \ifnum #1=1   \vcenter{\young}                       \fi
\ifnum #1=2
\vcenter{\vbox{\hbox to 0.425cm{\young\hss\young}}}           \fi
\ifnum #1=11
\vcenter{\vbox to 0.425cm{\young\vss\young}}                  \fi
\ifnum #1=3
\vcenter{\vbox{\hbox to 0.65cm{\young\hss\young\hss\young}}}  \fi
\ifnum #1=21
\vcenter{\vbox to 0.430cm{\hbox to 0.425cm{\young\hss\young}
\vss\hbox to 0.425cm{\young\hss}}}      \fi
\ifnum #1=111
\vcenter{\vbox to 0.65cm{\young\vss\young\vss\young}}         \fi
\ifnum #1=33
\vcenter{\vbox to 0.430cm{\hbox to 0.65cm{\young\hss\young\hss\young}
\vss\hbox to 0.65cm{\young\hss\young\hss\young}
}}      \fi
\ifnum #1=222
\vcenter{\vbox to 0.625cm{\hbox to 0.425cm{\young\hss\young}
\vss\hbox to 0.425cm{\young\hss\young}
\vss\hbox to 0.425cm{\young\hss\young}}}\fi
\ifnum #1=111111
\vcenter{\vbox to 1.3cm{\young\vss\young\vss\young
\vss\young\vss\young\vss\young}}\fi}
\begin{document}
\title{
      THE d'-DIBARYON IN THE NONRELATIVISTIC QUARK MODEL\\
}
\author{
      A. J. BUCHMANN, GEORG WAGNER, K. TSUSHIMA, AMAND FAESSLER  \\
{\it Institute for Theoretical Physics, University of T\"ubingen} \\
{\it D-72076 T\"ubingen, Germany}
\and
              L. YA. GLOZMAN \\
{\it Institute for Theoretical Physics, University of Graz} \\
{\it A-8010 Graz, Austria} }

\maketitle

\begin{abstract}
The narrow peak recently found in various pionic
double charge exchange (DCX) cross sections can be explained
by the assumption of a universal resonance at $2065$ MeV, called $d'$.
We calculate the mass of a six-quark system with
$J^P=0^-$, $T=0$ quantum numbers employing
a cluster model and a shell model basis to
diagonalize the nonrelativistic quark model Hamiltonian.
\end{abstract}
%
%
%
%
%
%
%

\noindent
\section {Introduction}

\noindent
Recently, a very narrow peak ($\Gamma_{medium} \approx 5$ MeV)
has been observed in the ($\pi^+$,$\pi^-$) double
charge exchange cross section at an incident pion energy
of $T_{\pi}=50$ MeV and forward angles $\Theta=5^o$ \cite{Bil93}.
Dedicated DCX experiments on a number of nuclei
ranging from $^{12}C $ to $^{56}Fe $
have shown that the position and width of this
peak is largely independent of the nuclear target.
Therefore, a fundamental two-nucleon or six-quark process seems
to be involved.
In fact, due to charge conservation the DCX reaction involves at least two
nucleons within the nucleus and is quite sensitive
to short-range $NN$ correlations \cite{Mil84}.
Conventional DCX calculations have so far been
unable to explain these experimental results \cite{Kag94}.
On the other hand, the assumption of a single narrow resonance, called $d'$,
with spin, parity $J=0^-$ and isospin $T=0$ and a resonance energy of
$M_{d'}=2065$ MeV works
extremely well in describing all available DCX data \cite{Bil93}.
The mass of the $d'$-resonance coincides
with $M_{d'}=2 M_N+ m_{\pi} +T_{\pi}$ \footnote{It has been suggested
that the $d'$-dibaryon is a $\pi NN$
resonance with isospin $T=2$ \cite{Gar94}. However, the $T=2$ assignment seems
to be in conflict with experimental information from other
sources \cite{Bil94}. }.
The small decay width of the $d'$
is naturally explained by its small mass
which is below any baryon-baryon
($NN^*, N^*N^*,...$) threshold and
by its quantum numbers J$^P$=0$^-$ and $T=0$ which prevent
a decay into the nucleon-nucleon channel.
\par
QCD does not exclude a rearrangement
of the six quarks into more exotic configurations such as a
diquark $(q^2)$ and a tetraquark ($q^4$) cluster.  In fact,
the bag-string model \cite{Mul78} predicts that
a diquark cluster with spin (isospin) $S=0$ ($T=0$)
and a tetraquark cluster with spin (isospin)  $S=1$ ($T=0$)
moving with relative angular momentum $L=1$
is energetically the most favorable configuration for a system
with $d'$ quantum numbers. Ref.\cite{Mul78} predicts a $d'$ mass of
$M_{d'} \approx 2100$ MeV. However, in these calculations
the quark exchange interactions between the clusters have been neglected.
Therefore, we  perform a calculation in the nonrelativistic
quark model (NRQM) in which the complications due to the Pauli principle
are properly taken into account.
We use two different sets of basis states
(i) a cluster model basis
using the Resonating Group Method (RGM),
(ii) a quark shell-model basis.
Furthermore, unlike the  bag-string model,
the NRQM calculation includes the nonperturbative $\pi$- and
$\sigma$-meson exchange interactions between quarks as required
by the spontaneously broken chiral symmetry of QCD.
\bigskip
\goodbreak
\noindent
\section {The nonrelativistic quark potential model}
\nobreak
\noindent
In the NRQM a system of $n$-quarks with equal masses $m_q=313$ MeV
is described by the Hamiltonian
\begin{equation}
\label{Ham}
H=
\sum_{i=1}^{n} \Bigl ( m_q+ {{\bf p}^{2}_{i}\over 2m_{q}} \Bigr )
-{{\bf P}^2\over n(2m_q)}
+ \sum_{i<j}^{n} V^{conf}({\bf r}_i,{\bf r}_j) +
\sum_{i<j}^{n} V^{res}({\bf r}_i,{\bf r}_j),
\end{equation}
where
${\bf r}_i$,
${\bf p}_i$
are the spatial and momentum coordinates of the i-th
quark, respectively and ${\bf P} $ is the total momentum of the
$n$-quark system. The exact removal of the kinetic energy
of the center of mass motion by the third term is crucial.
\par
The long-range confining force is generated by
a two-body harmonic oscillator confinement potential
\be
\label{conf}
V^{conf}({\bf r}_i,{\bf r}_j)=
-a_c \b{\lambda}_i\cdot \b{\lambda}_j
({\bf r }_i-{\bf r }_j)^2,
\ee
where
$\b{\lambda}_i$ is the Gell-Mann matrix of $SU(3)_{color}$ of the i-th quark.
The residual interaction models the relevant properties of QCD,
namely asymptotic freedom at short distances and spontaneous chiral
symmetry breaking which has important consequences
for the short- and intermediate-range interaction between quarks.
The one-gluon exchange potential \cite{deR75}
\be
\label{gluon}
V^{OGEP} ({\bf r}_i,{\bf r}_j)  =  {\alpha_{s}\over 4}
\b{\lambda}_i\cdot\b{\lambda}_j \Biggl \lbrace
{1\over r}-
{\pi\over m_q^2} \left ( 1+{2\over 3}
\b{\sigma}_{i}\cdot\b{\sigma}_{j} \right ) \delta({\bf r}) \Biggr \rbrace
\ee
provides an effective quark-quark interaction that
has the spin-color structure of QCD at short distances.
Here, ${\bf r}={\bf r}_i-{\bf r}_j$ and $\b{\sigma}_i$ is the usual
Pauli spin matrix.
\par
It is well known that the spontaneous breaking of chiral
symmetry by the physical vacuum is responsible for the constituent
quark mass generation \cite{Man84},
as well as for the appearence of almost massless
pseudoscalar Goldstone bosons and  massive scalar mesons
that couple to the constituent quarks.
In the NRQM, this is modeled by regularized
one-pion and one-sigma exchange potentials
between constituent quarks \cite{Obu90,Fer93}:
\bea
\label{Pion}
V^{OPEP}({\bf r}_i,{\bf r}_j) & = &
{g_{\pi q}^2\over 4 \pi }  {1\over  {4 m_q^2} }
{\Lambda^2\over {\Lambda^2-m_{\pi}^2}}
\b{\tau}_{i}\cdot \b{\tau}_{j}\,
\b{\sigma}_{i}\cdot\b{\nabla}_{r}\,
\b{\sigma}_{j}\cdot\b{\nabla}_{r}
\left ( {e^{-m_{\pi} r}\over r}- {e^{-\Lambda r}\over r} \right ), \nonumber \\
V^{OSEP}({\bf r}_i,{\bf r}_j) &= &-{g_{\sigma q}^2\over {4 \pi}}
{\Lambda^2\over {\Lambda^2-m_{\sigma}^2}}
\left ( {e^{-m_{\sigma} r}\over r}- {e^{-\Lambda r}\over r} \right ),
\eea
with
\be
\label{chiral}
{g^2_{\sigma q}\over {4 \pi}} = {g^2_{\pi q}\over {4 \pi}}, \qquad
m_{\sigma}^2 =(2 m_q)^2+ m_{\pi}^2,  \qquad
\Lambda_{\pi} =  {\Lambda_{\sigma}}\equiv \Lambda.
\ee
The $\pi q$ coupling constant is simply
related to the $\pi N$ coupling constant \cite{Obu90,Fer93}.
We use $f^2_{\pi N} /4 \pi=0.0749$ \cite{Ber90}.
The $\pi q$ cut-off mass $\Lambda$ describes the extended pion-quark vertex.
Here, $\Lambda=4.2$ fm$^{-1}$ which results in a soft $\pi N$ form
factor \cite{Buc91}.
\par
We take a harmonic oscillator wave function
for the i-th quark inside the nucleon and $\Delta$
\be
\varphi^{N (\Delta)}({\bf r}_i)
=\left ({1\over {\pi b^2}} \right )^{3/4}
\exp \left (-{1\over 2b^2} {\bf r}_i^2  \right ),
\ee
where $b$ is the harmonic oscillator constant.
%
%
\goodbreak
\begin{table}[htb]
\caption[Quark model parameters]{Quark model parameters.
Set I: with regularized $\pi$- and $\sigma$-meson exchange potentials.
The $\sigma$ parameters are fixed by eq.(\ref{chiral}).
Set II: without $\pi$- and $\sigma$-exchange potentials \cite{Buc91}.
Set III: as set I but with reduced confinement strength (see sect. 4).
 }
\begin{center}
\begin{tabular}[h]{|r|r|r|r|r|r|r|} \hline
  Set     & $b\,$ [fm]      &  $\alpha_s$      &  $a_c$ [MeV\,fm$^{-2}]$ &
    $m_{\sigma}$ [MeV] & $g^2_{\sigma}/( 4 \pi)$ & $ \Lambda$ [fm$^{-1}] $
          \\ \hline
 I    &   0.595     &   0.958   & 13.66   &    626     &   0.554
& 4.2   \\ \hline
 II  & 0.603  &     1.540  & 24.94  &   ---    &   ---  &     ---
\\  \hline
 III    &   0.595     &   0.958   & 5.00   &    626     &   0.554
& 4.2   \\ \hline
\end{tabular}
\end{center}
\end{table}
\noindent
We determine the parameters
$a_c$, $\alpha_s$, and $b$ by requiring that
the $N(939)$ and $\Delta (1232) $ masses are reproduced
and that the nucleon mass is stable with respect to variations
in $b$.
\be
\label{constraints}
M_{N}(b)=3m_q=939 MeV, \qquad
M_{\Delta}-M_{N}=293 MeV, \qquad
{\partial M_{N}(b) \over \partial b}= 0.
\ee

\section{Six-Quark Models for the d'-Dibaryon}

\noindent
In the bag-string model, dibaryons are described as rotating
strings with colored quark clusters at the ends \cite{Mul78}.
This model leads to a linear Regge trajectory of excited
states. It predicts that the lowest $L=1$ excited state
is obtained for a diquark-tetraquark configuration at around 2100 MeV
\cite{Mul78}.

\begin{figure}[htb]
\label{Fig.1}
$$\mbox{
\epsfxsize 10.0 true cm
\epsfysize 5.5 true cm
\setbox0= \vbox{
\hbox { \centerline{
\epsfbox{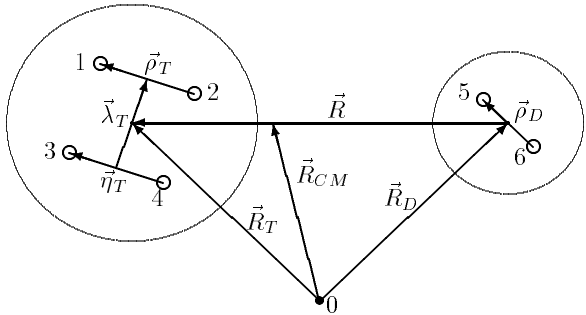}
}  } 
} 
\box0
} $$
\vspace{0.1cm}
\caption[The d'-dibaryon in the quark cluster model]
{The d'-dibaryon in the $q^4$-$q^2$  quark cluster model.}
\end{figure}
%
%
The drawback of this approach is that it does not respect
the Pauli principle. Only the quarks within
the individual clusters are antisymmetrized but not the quarks
belonging to  different clusters. This is a good approximation
for high angular momentum states because in this case
the system is fairly elongated and the probability of
cluster overlap is small. On the other hand, for a low lying
$L=1$ excitation, such as the $d'$  one expects a considerable amount
of quark exchange between the two clusters.
\begin{table}[htb]
\caption{The mass ($M_{d'}$) and size ($b_6$) of the $d'$
in the quark cluster model. The masses of the diquark and tetraquark
are also given.}
\begin{center}
\begin{tabular}{|c||c|c|c|c|}\hline
 & $M_{2q}$ & $M_{4q}$ & $M_{d'}$ & $b_6$\\
Set I    & 645 & 1455 & 2440&  0.76  \\
Set II   & 637  & 1501 & 2634 & 0.70  \\
Set III   & 621  & 1309  & 2111  & 0.95      \\
\hline
\end{tabular}
\end{center}
\end{table}
The confining forces between the colored quarks
prevent large separations of the clusters and the typical size of
such a system is expected to be about 1 fm.
{}From our experience with the $NN$ system we know that the Pauli
principle plays an important role at such distances \cite{Yam91}.
\vspace{0.5cm}

\noindent
{\it 3.1~ The Quark Cluster Model of the $d'$-Dibaryon }

\vspace{0.5cm}
\nobreak
\noindent
In this model, the $d'$ is described as
a nonrelativistic six-quark system in which the quarks
interact via the two-body potentials
of eq.(\ref{Ham}). Tensor and spin-orbit interactions have been omitted
since it has previously  been shown
that they give a negligible
contribution to the $d'$ mass \cite{Glo94,Wag95}.
The six-quark wave function is expanded into the cluster
basis
\bea
\label{rgmwf}
\mid \Psi_{d'}^{J=0,T=0}>
& = & {\cal A} {\Bigl \vert}
\Biggl [ \Bigl [ \Phi_{T}^{S_T=1,T_T=0}
(\b{\rho}_{T},\b{\lambda}_{T},\b{\eta}_T)
\times \ \Y{211}_C
\nonumber \\
& & \otimes\Phi_{D}^{S_D=0,T_D=0} (\b{\rho}_{D}) \times \Y{11}_C
\Bigr ]^{S=1,T=0}
\otimes\chi_{L=1}({\bf R}) \Biggr ]^{J=0,T=0} \ \Y{222}_C \Bigr >,
\eea
where
$\Phi_T^{S_T=1,T_T=0}(\b{\rho}_{T},\b{\lambda}_{T},\b{\eta}_T)$
and
$\Phi_B^{S_D=0,T_D=0}(\b{\rho}_{D})$
are the internal wave functions of the tetraquark (T)
and diquark (D) clusters, respectively and
$\chi_{L=1}({\bf R})$ is the wave function of the relative motion of the
two clusters.
We use the same harmonic oscillator parameter for
the internal and relative motion wave functions.
The Young diagrams in eq.(\ref{rgmwf}) show
that two color triplet clusters $[211]_C$ and $[11]_C$
are coupled to a $[222]_C$ color-singlet six-quark state.
Furthermore, they show that the tetraquark and $d'$ wave function
are not fully antisymmetric but have mixed symmetry in color space.
This nonfactorizability of the color space
considerably complicates the calculation.
%
%
\begin{figure}[htb]
\label{Fig.2}
$$\mbox{
\epsfxsize 11.5 true cm
\epsfysize 6.5 true cm
\setbox0= \vbox{
\hbox { \centerline{
\epsfbox{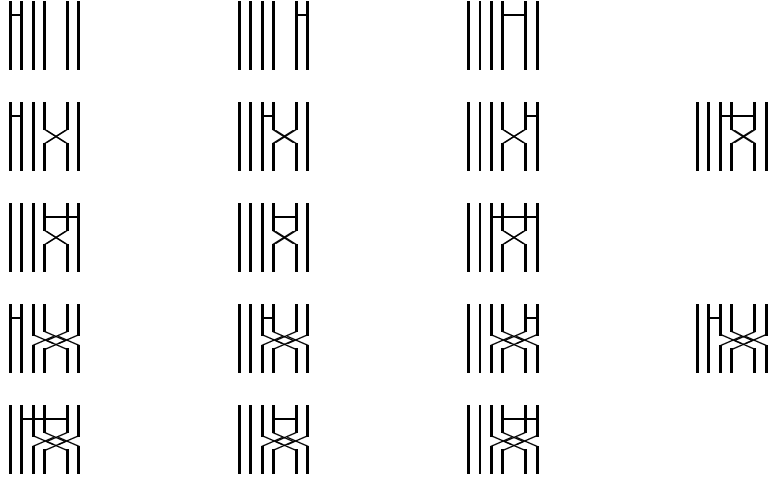}
}  } 
} 
\box0
} $$
\vspace{0.2cm}
\caption[Potential matrix elements]
{The direct, one-quark and two-quark
exchange diagrams that have to be evaluated for each two-body potential.
The horizontal bars indicate the confinement, the
one-gluon, one-pion, or one-sigma
exchange interactions in eq.( \ref{Ham}).}
\end{figure}
%
\par
The advantage of the cluster model is that it provides
a continuous transition from the $q^6$ six-quark state to the
$q^4-q^2$ clusterized state by smoothly going through all intermediate
configurations. There is no rigid and artificial boundary between
these extreme configurations;
everything is contained in one and the same RGM wave function.
This important property is a direct consequence of the Pauli principle
on the quark level, which is ensured by the
antisymmetrizer ${\cal A}$
\be
\label{ant}
{\cal A}= 1 - 8P_{46}^{OSTC} +6P_{35}^{OSTC} P_{46}^{OSTC},
\ee
where $P_{ij}^{OSTC}$ is the permutation operator of
the i-th and j-th quark in orbital (O)
spin-isospin (ST)  and color space (C), respectively.
The direct, as well as the one- and
two-quark exchange contributions for the two-body potential of
eq.(\ref{Ham}) are depicted in fig.2.
The solution for the unknown relative wave function $\chi_L({\bf R})$
and the unknown eigenenergy is obtained from the variational principle
\be
\delta\left[{ \langle\Psi_{d'}\vert H-E\vert \Psi_{d'} \rangle
\over\langle\Psi_{d'}\vert \Psi_{d'}\rangle  }\right ]=0,
\ee
where the variation is with respect to the relative wave function
$\chi_L ({\bf R}) $. The results for the energy (mass) of the $d'$
as well as for the harmonic oscillator parameter
$b_6$ which minimizes the $d'$ mass are shown in table 2 for the parameter
sets of table 1.
\vspace{0.5cm}

\noindent
{\it 3.2~ Shell-Model Calculation for a J$^P$=0$^-$, T=0 six-quark system}

\vspace{0.5cm}
\noindent
Next, we calculate the mass of the
$d'$-dibaryon in the translationally invariant
shell-model (TISM) \cite{Glo94,Wag95}.
Due to the negative parity of the $d'$, only an odd
number of oscillator quanta
$N=1,3,5,...$ is allowed.
There is only one  $N=1$ state
which is compatible with $J^p=0^-,T=0$
\be
\label{smgs}
\mid \Psi_{d'_{g.s.}}> =
\mid N=1, [51]_O, (\lambda\mu)=(10), L=1, S=1, T=0, [321]_{ST}>.
\ee
For an unambigious classification of TISM states one has
to specify the number of internal
excitation quanta $N$, the Elliot symbol $(\lambda\mu)$, the Young
pattern $[f]_O$ of the spatial permutational
$S_6$-symmetry, further the total orbital angular momentum $L$,
total spin $S$ and total isospin $T$ of the system.
The specification of the intermediate $SU(4)_{ST}$ symmetry is
necessary because in general, the same symmetry in $STC$ space can be
obtained from several states with different intermediate $ST$ symmetries.
The mass of the $d'$ is then given in first order perturbation theory
by the expectation value of the Hamiltonian between the
lowest harmonic oscillator state
of eq.({\ref{smgs})
\be
\label{smm}
M_{d'}(b_6) = <\Psi_{d'_{g.s.}} \mid  H  \mid \Psi_{d'_{g.s.}}>.
\ee
In order to estimate the effect of configuration mixing
with excited shell model states we include
in addition ten $N=3$ states with orbital $[42]_O$ symmetry \cite{Wag95}.
In this case also the $[51]_{ST}$, $[411]_{ST}$, $[33]_{ST}$,
$[321]_{ST}$, and $[2211]_{ST}$ $S_6$ permutational symmetries are allowed.
\par
With  fixed parameters of the quark-quark interaction
determined from eq.(\ref{constraints}) we minimize
the $d'$ mass with respect to the harmonic oscillator parameter $b_6$
in the six-quark wave function. Note, that the harmonic oscillator
parameter of the  single baryon ($b$) and
the $d'$ ($b_6$) wave function are different.
The value of $b_6$ which minimizes the $d'$ mass
is a measure of the size of the system and is
also given in table 3.

\begin{table}[htb]
\caption{The mass ($M_{d'}$) and size ($b_6$) of the $d'$
in the six-quark shell model without and with configuration mixing.}
\begin{center}
\begin{tabular}{|c||c|c||c|c|}\hline
& \multicolumn{2}{c||}{$N=1$} & \multicolumn{2}{c|}{$N=1$ $\&$ $N=3$ } \\
\hline
 & $M^{(N=1)}_{d'}$ [MeV] & $b_6$ [fm] & $M_{d'}$ [MeV] & $b_6$ [fm] \\
\hline
Set I    &  2484  & 0.78  & 2413   & 0.78  \\
Set II   &  2636  & 0.72    & 2553   & 0.73 \\
Set III  &  2112  & 0.95   & 2063   & 0.96  \\
\hline
\end{tabular}
\end{center}
\end{table}

\section{ Discussion and Summary}

\noindent
As is evident from tables 2 and 3 the $d'$ mass of the cluster model
is lower than the single $N=1$ shell-model mass of eq.(\ref{smm})
but higher than the shell-model result with
configuration mixing. Note that the $d'$ mass calculated with set II
(without pion and sigma-exchange between quarks)
is some 150-200 MeV higher than the result with chiral interactions (set I).
In any case, the calculated mass is about 350 MeV higher than the
value required by experiment. However, the confinement strength $a_c$
in the three-quark and six-quark system need not be the same.
If we assume (set III) that $a_c$ in the six-quark system is weaker
than in the nucleon one obtains considerably
smaller values for $M_{d'}$. This assumption is supported by the
harmonic oscillator relation for $a_c$
\be
\label{confstr}
a_c \propto {1\over m_q b^4} {1\over N}
\ee
which is inversely proportional to the number of quarks $N$ in the system.
A weaker confinement strength is also expected due to the larger
hadronic size of the $d'$ ($b_6$) as compared to the
hadronic size of the nucleon $(b)$.
Set III differs from Set I of table 1 only in the strength of
the parameter $a_c$ for which we take the value
$a_c=5.0$ MeV/fm$^2$ in the six-quark calculation.
Finally, both calculations
give similar results for the
$d'$ mass and for its size.
Let us briefly discuss the reasons for this.
The outer product of the orbital
$[4]_O$ (tetraquark) and $[2]_O$ (diquark)
symmetries gives the following six-quark symmetries
\be
[4]_O \otimes [2]_O   =  [42]_O \ \   \oplus \ \ [51]_O \ \  \oplus \ \ [6]_O.
\ee
With the exception of the $[6]_O$ symmetry which is incompatible with
$d'$ quantum numbers these are
also included in the enlarged $N=3$ shell-model basis \cite{Wag95}.
Analogously, the outer product of the two clusters in spin-isospin space
leads to
\be
\label{STSYM}
[31]_{ST} \otimes [2]_{ST}   =  [51]_{ST}  \ \ \oplus \ \ [42]_{ST} \ \
\oplus \ \ [33]_{ST}  \ \ \oplus \ \  [411]_{ST} \ \  \oplus \ \  [321]_{ST}.
\ee
Comparison with eq.(10) in ref.\cite{Glo94} shows
that the $q^4-q^2$ cluster model wave function comprises
the same $S_6$-symmetries in orbital and spin-isospin space
(with the exception of the $[2211]_{ST}$ symmetry) as our
enlarged shell model basis. Thus the trial function space spanned by
both sets of basis functions is not very different.
\par
In summary, we have calculated the mass of a $J^P=0^-$ $T=0$
six-quark system in the NRQM using two different assumptions
for the spatial distribution of the six quarks. The parameters
have been determined from the constraints of eq.(\ref{constraints}).
As in our previous works \cite{Glo94,Wag95}
our results are typically 300-400 MeV above the
required resonance energy. However, for a weaker
confinement strength $a_c$ in the six-quark system as suggested by
eq.(\ref{confstr}) we find a mass for the $d'$ that is considerably smaller.
The assumption of a weaker confinement strength in the six-quark system
does not affect previous results of the model in the $B=2$ sector such as
$NN$ scattering phase shifts or deuteron electromagnetic form factors
which are completely insensitive to the model and strength
of confinement \cite{Shi89}.

\end{document}